\documentclass[prl,superscriptaddress,twocolumn]{revtex4-1}

\usepackage{graphicx}
\usepackage{amsmath}
\usepackage{sidecap,textcomp}
\usepackage{color}
\usepackage[normalem]{ulem}
\usepackage[utf8]{inputenc}

\begin{document}
\newcommand{\sd}[1]{\textcolor{red}{#1}}
\newcommand{\hrt}[1]{\textcolor{blue}{#1}}

\title{$\chi^{(2)}$ mid-infrared frequency comb generation and stabilization with few-cycle pulses}

\author{Alexander J. Lind}
\thanks{These authors contributed equally to the manuscript.}
\affiliation{Time and Frequency Division, National Institute of Standards and Technology, 325 Broadway, Boulder, Colorado 80305, USA}
\affiliation{Department of Physics, University of Colorado, 2000 Colorado Ave., Boulder, Colorado 80309, USA}
\author{Abijith Kowligy}
\thanks{These authors contributed equally to the manuscript.}
\affiliation{Time and Frequency Division, National Institute of Standards and Technology, 325 Broadway, Boulder, Colorado 80305, USA}
\affiliation{Department of Physics, University of Colorado, 2000 Colorado Ave., Boulder, Colorado 80309, USA}
\author{Henry Timmers}
\thanks{These authors contributed equally to the manuscript.}
\affiliation{Time and Frequency Division, National Institute of Standards and Technology, 325 Broadway, Boulder, Colorado 80305, USA}
\author{Flavio C. Cruz}
\affiliation{Time and Frequency Division, National Institute of Standards and Technology, 325 Broadway, Boulder, Colorado 80305, USA}
\affiliation{Instituto de Fisica Gleb Wataghin, Universidade Estadual de Campinas, Campinas, SP, 13083-859, Brazil}
\author{Nima Nader}
\affiliation{Applied Physics Division, National Institute of Standards and Technology, 325 Broadway, Boulder, Colorado 80305, USA}
\author{Myles C. Silfies}
\author{Thomas K. Allison}
\affiliation{Stony Brook University, Stony Brook, New York 11794, USA}
\author{Scott A. Diddams}
\email{scott.diddams@nist.gov}
\affiliation{Time and Frequency Division, National Institute of Standards and Technology, 325 Broadway, Boulder, Colorado 80305, USA}
\affiliation{Department of Physics, University of Colorado, 2000 Colorado Ave., Boulder, Colorado 80309, USA}

\begin{abstract}
Mid-infrared laser frequency combs are compelling sources for precise and sensitive metrology with applications in molecular spectroscopy and spectro-imaging. The infrared atmospheric window between 3\textendash{}5.5 \textmu m in particular provides vital information regarding molecular composition. Using a robust, fiber-optic source of few-cycle pulses in the near-infrared, we experimentally demonstrate ultra-broad bandwidth nonlinear phenomena including harmonic and difference frequency generation in a single pass through periodically poled lithium niobate (PPLN). These $\chi^{(2)}$ nonlinear optical processes result in the generation of frequency combs across the mid-infrared atmospheric window which we employ for dual-comb spectroscopy of acetone and carbonyl sulfide with resolution as high as 0.003 cm$^{-1}$. Moreover, cascaded $\chi^{(2)}$ nonlinearities in the same PPLN directly provide the carrier-envelope offset frequency of the near-infrared driving pulse train in a compact geometry. 
\end{abstract}

\maketitle
%




Coherent laser sources in the mid-infrared (MIR, 3\textendash{}25 \textmu m) have long been recognized as important tools for both fundamental and applied spectroscopy and sensing. Recently, significant interest has focused on developing laser frequency combs in the MIR spectral region \cite{schliesser_mid-infrared_2012}.  Of many promising applications, molecular spectroscopy using optical frequency combs benefits from a unique combination of high spectral resolution and broad bandwidth. This is particularly useful for the simultaneous measurement of spectral absorption fingerprints for a wide range of molecular compounds. The infrared atmospheric window between 3\textendash{}5.5 \textmu m exhibits reduced atmospheric attenuation while demonstrating strong absorption coefficients for greenhouse gases and pollutants such as methane, ethane, carbon dioxide, and formaldehyde \cite{baumann_spectroscopy_2011, mulrooney_detection_2007,golston_lightweight_2017,loh_absorption_2017,zhu_observing_2016,lancaster_difference-frequency-based_2000,bjork_direct_2016,fleisher_mid-infrared_2014}, making this spectral range useful for climate research and atmospheric monitoring. Further, the same spectral window contains important molecular structure information pertaining to the C-H and O-H functional groups which can be used in the characterization of complex biochemical molecules \cite{clemens_vibrational_2014,maquelin_identification_2002} and spectro-imaging of biological samples \cite{bhargava_infrared_2012,walsh_label-free_2012}. 

Based on these motivations, multiple approaches to MIR frequency comb generation have been pursued. Examples include optical parametric oscillators (OPOs) \cite{leindecker_octave-spanning_2012,adler_phase-stabilized_2009,balskus_mid-infrared_2015,muraviev_massively_2018,jin_femtosecond_2015}, supercontinuum generation \cite{hickstein_ultrabroadband_2017,nader_versatile_2018,guo_mid-infrared_2018,lau_octave-spanning_2014}, difference frequency generation (DFG) \cite{cruz_mid-infrared_2015,erny_mid-infrared_2007,ycas_high-coherence_2018,yan_mid-infrared_2017}, direct generation with quantum cascade lasers (QCL) \cite{hugi_mid-infrared_2012, piccardo_widely_2018,barbieri_coherent_2011}, mode-locked fiber lasers \cite{hu_ultrafast_2015,henderson-sapir_mid-infrared_2014,duval_femtosecond_2015}, and microresonator frequency combs \cite{yu_mode-locked_2016}. 
Despite significant progress, many of these frequency comb sources require additional resonant cavities (OPOs) or careful spatio-temporal alignment of two femtosecond pulses (DFG) that increases complexity.  Others lack absolute frequency calibration and have large mode-spacings (microresonator combs and QCLs) that are mis-matched to the spectroscopy of small molecules.


In this Letter, we introduce a simple and powerful method for generating frequency combs across the MIR atmospheric window using intra-pulse DFG, a form of optical rectification driven by few-cycle pulses \cite{timmers_molecular_2018,pupeza_high-power_2015,kowligy_mid-infrared_2018}. This method avoids the strict requirement on spatio-temporal overlap, which reduces complexity and eliminates a major source of intensity noise, and is achieved in a single-pass, bypassing the need for resonant cavities. The process is driven by robust and commercially-available fiber laser technology starting in the 1.55 \textmu m telecom region and only requires a commonplace periodically poled lithium niobate (PPLN) crystal.
Beyond producing a MIR frequency comb, the output of the PPLN directly provides the carrier-envelope offset frequency of the 1.55 \textmu m Er:fiber mode-locked laser, which we use to fully stabilize the near infrared (NIR) comb spanning 1-2 \textmu m. We employ the MIR comb for dual comb spectroscopy (DCS) on acetone and carbonyl sulfide (OCS) in the MIR, highlighting the excellent signal-to-noise ratio (SNR) and capability to accurately resolve both broad- and narrow-band spectral features. 



\begin{figure}
\centering
\includegraphics[width=\linewidth]{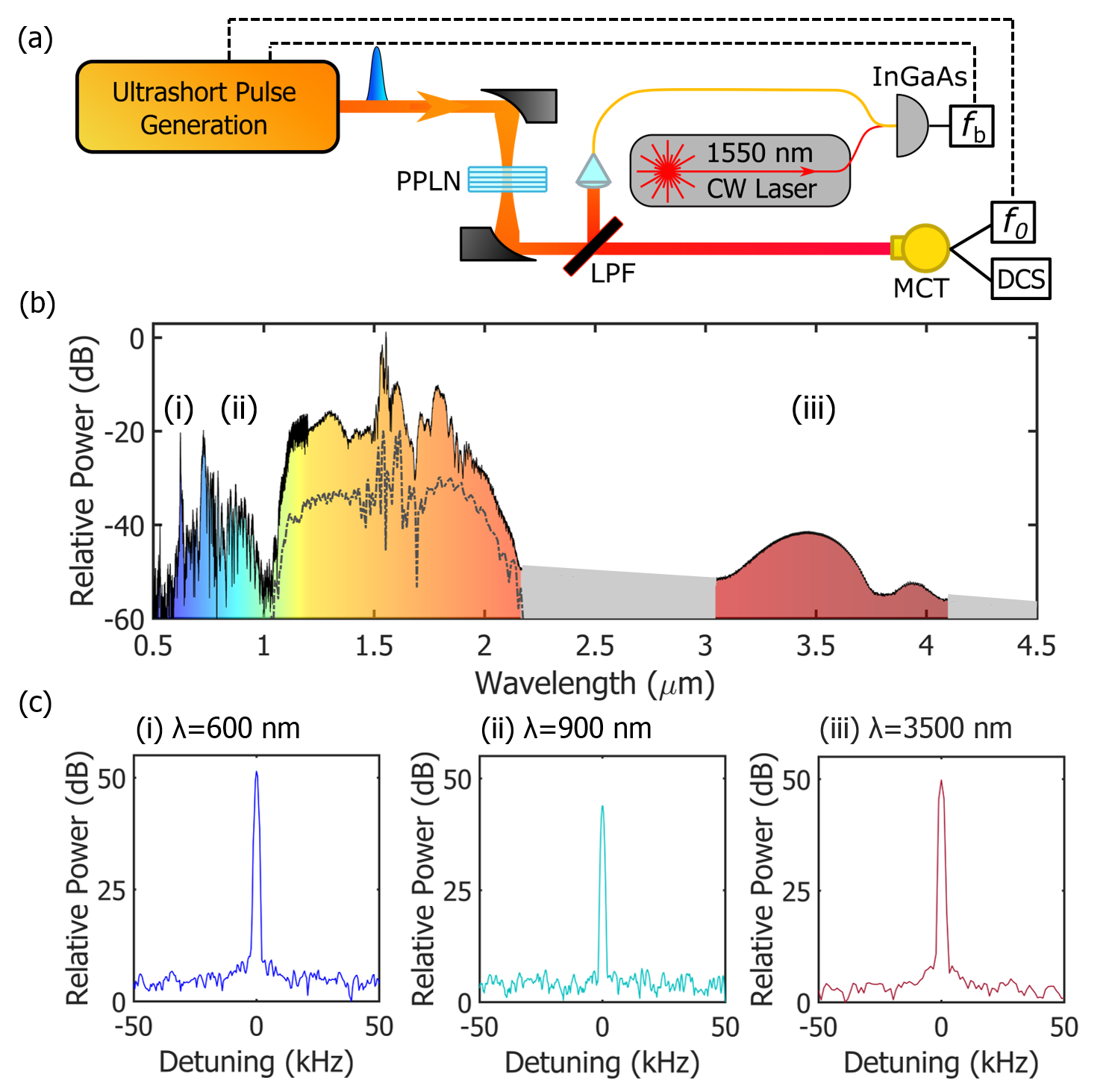}
\caption{Experimental overview and offset frequency detection. (a) Ultrashort pulses (10 fs) are generated from an Er:fiber oscillator at 100 MHz. The light is focused into periodically poled lithium niobate (PPLN) crystals. A long-pass filter (LPF) allows the MIR light to pass and focus onto a MCT detector. Additionally, the offset frequency ($f_0$) is detected here. The light reflected from the LPF is coupled into optical fiber to heterodyne with a cavity-stabilized 1.55 \textmu m laser, which is used to fully stabilize the comb. (b) A full optical spectrum before (gray dashed line, vertically offset for clarity) and after (shaded line, gray regions indicate noise floor) propagation through a PPLN crystal. (c) Offset frequency beat notes taken at three different spectral locations with three different detectors: (i) 600 nm with a Si photodetector; (ii) 900 nm with an InGaAs photodetector; and (iii) 3500 nm with a liquid nitrogen-cooled MCT photodetector. All beat notes are recorded with 1 kHz resolution bandwidth.}
\label{Fig:expSetup}
\end{figure}

The experimental setup is shown in Fig \ref{Fig:expSetup}(a). With commercial fiber components, we amplify and spectrally broaden the output of a 100 MHz Er:fiber oscillator to generate 3 nJ few-cycle pulses (10.6 fs, 2.1 cycles) centered at 1.55 \textmu m \cite{timmers_molecular_2018}. We focus the few-cycle pulse with a 25 mm focal length silver-coated off-axis parabolic (OAP) mirror into 1 mm-long PPLN crystal, achieving a minimum beam diameter of 16 \textmu m. The broadband output of the PPLN crystal is collected with another OAP mirror, optically filtered, and measured on a commercial Fourier transform spectrometer (FTS) or sent to a liquid nitrogen-cooled mercury cadmium telluride (MCT) detector. 

The full optical spectrum after the PPLN crystal is shown in Fig \ref{Fig:expSetup}(b). The spectrum of the input few-cycle pulse spans 1\textendash 2 \textmu m (dashed curve in Fig. \ref{Fig:expSetup}(b)). Below 1 \textmu m, the sharp spectral features come from higher-order quasi phase-matched (QPM) harmonic generation from the input spectrum. The MIR light arises from difference-frequency mixing within the input spectrum.

The modes of a frequency comb are defined through two radio frequency parameters, the repetition rate ($f_{rep}$) and the offset frequency ($f_0$), such that a given mode, $n$, has a well-defined optical frequency, $\nu_n = nf_{rep} + f_0$. For DFG that occurs strictly within the original input pulse, the resulting MIR light will be ``offset-free'', meaning the offset frequency of the comb will subtract out in the DFG process. However, due to the high peak intensities within the crystal, cascaded $\chi^{(2)}$ nonlinear processes give rise to additional comb modes throughout the spectrum with which we observe $f_0$ in heterodyne beats at multiple wavelengths, as shown in Fig \ref{Fig:expSetup}(c). For example, at 600 nm, we observe the $2f$\textendash{}$3f$ interference, from cascaded QPM 1.8 \textmu m tripling and 1.2 \textmu m (third-order) doubling at a beat frequency, $f_b$,
\begin{equation}
f_b = 3\times(n f_{rep} + f_0) - 2\times(\frac{3}{2}n f_{rep}+f_0) = 3 f_0 - 2 f_0 = f_0.
\end{equation}
At 900 nm, a $f$\textendash{}$2f$ interference results from QPM 1.8 \textmu m frequency doubling beating with native comb light. At 3500 nm, we observe an ``$f$\textendash{}$0$'' beat note between the ``offset-free'' DFG comb, and DFG between doubled light from 2 \textmu m  (with a factor of $2f_0$) and the original 1.5 \textmu m comb (containing a single $f_0$) \cite{fuji_monolithic_2005}. The two MIR combs heterodyne together on the MCT photodetector to generate an offset frequency tone in the RF spectrum.

We implement optical frequency control at 1.55 \textmu m, which ensures a high degree of mutual coherence necessary for applications like DCS. A NIR comb tooth heterodyned against a stable 1.55 \textmu m laser provides one lock parameter, and an offset frequency beat provides the second. An additional external $f$\textendash{}$2f$ interferometer  branch is usually required to detect an offset frequency beat note; however, in this configuration, we bypass the need for such a branch by utilizing the offset frequencies directly detected throughout the spectrum. For the remaining experiments in this paper, we use the beat note from the MCT detector used for MIR spectroscopy to lock the offset frequency of our combs. The integrated phase noise on the offset frequency is 115 mrad, integrated from 50 Hz to 2 MHz, and comparable to other methods of offset frequency stabilization. This demonstrates that using the $f_0$ in the MIR results in similar performance to a traditional $f$-$2f$ lock, allowing for simultaneous comb stabilization and MIR spectroscopy from the same detector.



\begin{figure}
\centering
\includegraphics[width=\linewidth]{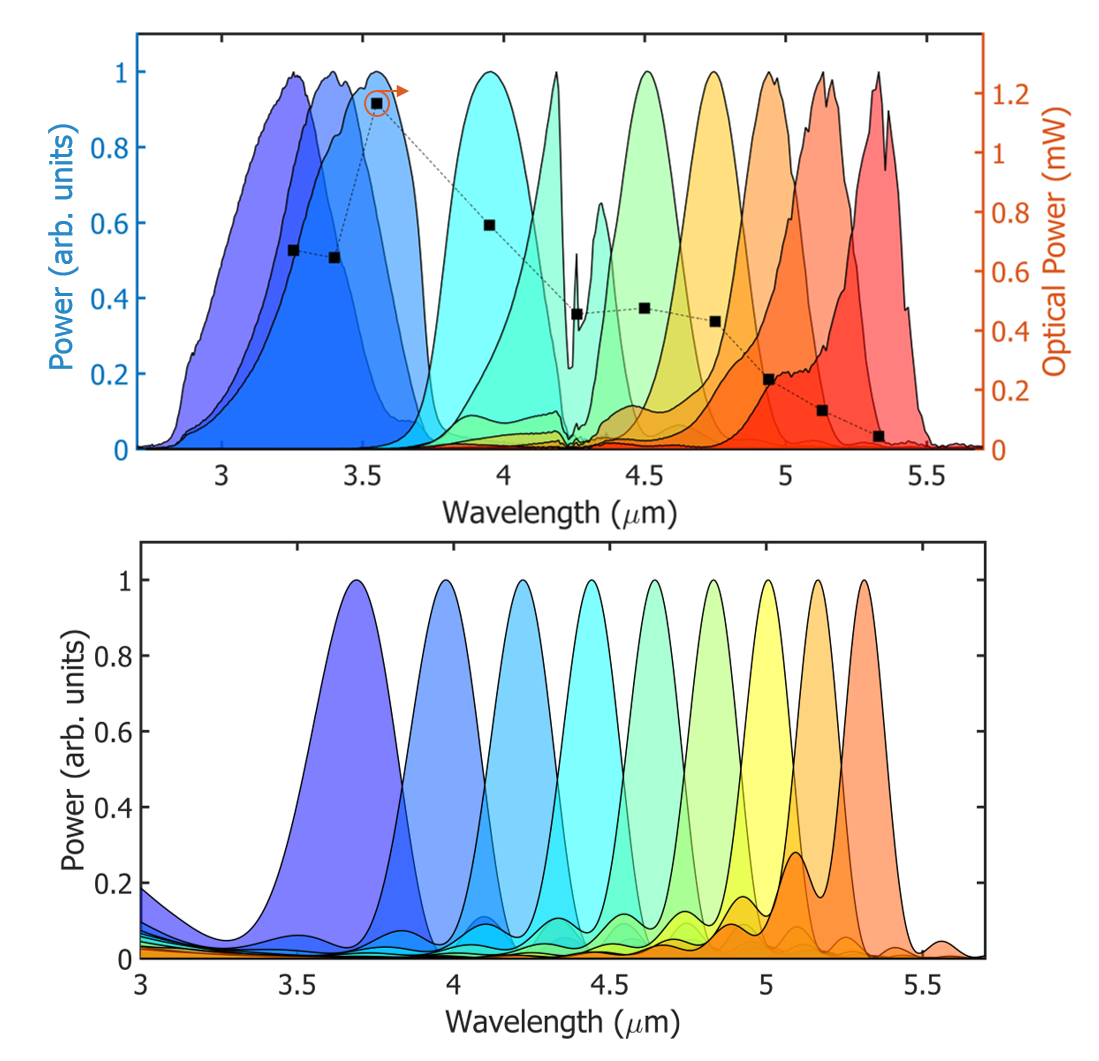}
\caption{MIR Spectra. (a) Experimentally measured intra-pulse DFG spectra. Measured with a commercial FTS (4 cm$^{-1}$/120 GHz resolution). All spectra are normalized, with measured power indicated by the square markers. Atmospheric CO$_2$ absorption is visible in the spectrum centered around 4.2 \textmu m. (b) Predicted MIR spectra, also normalized.}
\label{Fig:tunability}
\end{figure}

The PPLN crystal contains a set of discrete poling periods, ranging from $\Lambda = $ 24.1\textendash{}35.6 \textmu m in steps of approximately 0.6 \textmu m. The longer (shorter) poling periods provide QPM for DFG into shorter (longer) wavelengths in the MIR. By tuning across these poling periods, MIR light spans 3 - 5.5 \textmu m (Fig \ref{Fig:tunability}(a)). The maximum power measured in the MIR spectra is found to be 1.3 mW.  However, on average these spectra yield approximately 500 \textmu W, which is sufficient to saturate high-speed, low-noise MCT detectors in lab-scale linear spectroscopy measurements.

To develop physical insight into the nonlinear dynamics driven by the ultrashort pulse, we model $\chi^{(2)}$ processes using a nonlinear envelope equation. In contrast to conventional DFG which can be modeled as a set of coupled equations, the few-cycle nature of the pump pulse requires a broadband equation which does not break down at large bandwidths \cite{brabec_nonlinear_1997}. The governing equation is
\begin{align}
\frac{\partial A}{\partial z}+ &i\hat{D}A(z,t) = i\left(1 + \frac{i}{\omega_0}\frac{\partial}{\partial t}\right)\times\\\nonumber
&\chi(z)\left(A^2e^{-i\phi(z,t)} + |A|^2e^{i\phi(z,t)}\right),
\end{align}
where $A$ is the envelope of the electric field, $\hat{D} = \sum_{j=2}\frac{1}{j!}k_j(i\frac{\partial}{\partial t})^j$ is the dispersion operator, $\chi(z) = \chi^{(2)}(z)\omega_0^2/4\beta_0c^2$ accounts for periodic poling and the nonlinear susceptibility, $\phi(z,t) = \omega_0 t-(\beta_0 - \beta_1 \omega_0)z$, and $\beta_j$ are the coefficients of a Taylor expansion of $\beta(\omega)=\omega n(\omega)/c$, the propagation constant, about $\omega_0$ \cite{conforti_nonlinear_2010}. In the second line of the equation, the first term accounts for harmonic and sum frequency generation, while the second term is responsible for difference frequency generation or optical rectification. We numerically integrate Eq. (2) to model the propagation of a few-cycle pulse centered at 1.55 \textmu{}m through a 1-mm PPLN crystal with results summarized in Fig \ref{Fig:tunability}(b). The locations of the spectral peaks from the model agree well with experimental data, as do the relative bandwidths, i.e. the light generated closer to 3 \textmu m has a larger bandwidth than the light generated near 5.5 \textmu m. 

\begin{figure}
\centering
\includegraphics[width=\linewidth]{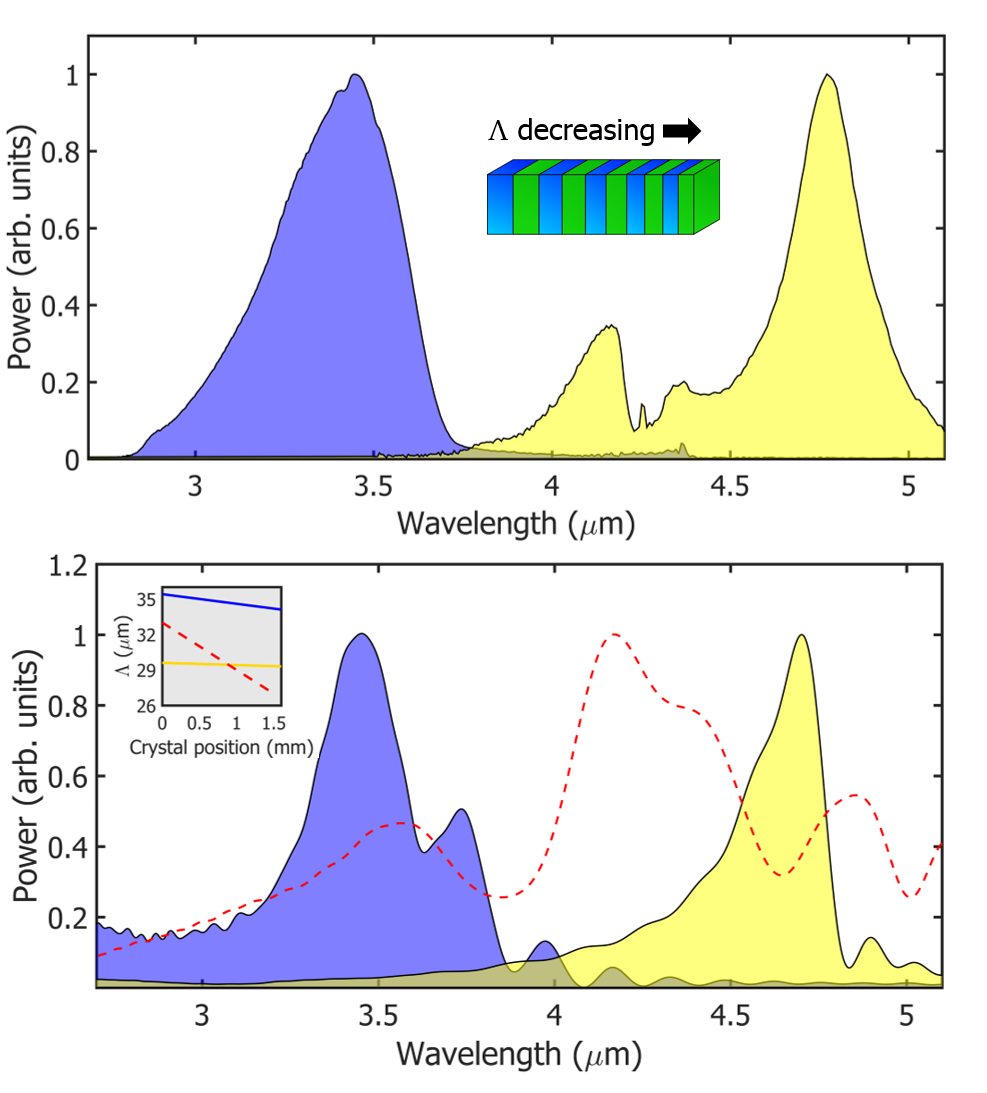}
\caption{Broad bandwidth MIR generation. (a) Using a fanout PPLN crystal tilted at an angle, we generate broader bandwith MIR light by propagating through several poling periods, effectively adding a chirp to the poling periods. CO$_2$ absorption is also visible in the yellow spectrum. Powers for the two spectra are 1.3 mW (purple) and 360 \textmu W (yellow). (b) Modeling of the experimental spectra. The chirp profiles are shown in the inset. The red dashed line is a modeled profile which covers this full spectral region. With this faster chirp and higher order dispersion compensation, we could generate up to 6 mW across this band. The chirp rate for each spectrum is: 0.8 \textmu m/mm (purple), 0.2 \textmu m/mm (yellow), and 4\textmu m/mm (red dashed).}
\label{Fig:chirp}
\end{figure}

The bandwidth generated (300\textendash{}500 nm) in the MIR is limited by group velocity dispersion elongating the pulse (112 fs$^2$/mm at 1.55 \textmu{}m) and the phasematching bandwidth for a single poling period. Instead, by using crystals with a chirped poling period (aperiodic poling), we significantly increase the QPM bandwidth \cite{charbonneau-lefort_competing_2008}. To implement a chirped poling period, we use a fanout PPLN crystal tilted at approximately 45 degrees, such that the pulse experiences different poling periods as it passes through the crystal (Fig \ref{Fig:chirp}(a)). Through these chirped grating periods, we generate bandwidths $>$1000 nm. The chirp, and thus the generated bandwidth, is limited by the crystal angle in the current implementation. 

Using Eq. (2), and grating chirp profiles shown in the inset of Fig \ref{Fig:chirp}(b), we accurately predict the general features of the experimental data (Fig \ref{Fig:chirp}(b)). With an optimized PPLN chirp profile (Fig \ref{Fig:chirp}(b), red dashed line) our modeling shows it would be possible to cover the entire 3-5 \textmu{]m spectral region in a single pass, allowing for the simultaneous probing of many molecular species \cite{muraviev_massively_2018}.


\begin{figure}[h!]
\centering
\includegraphics[scale = 0.50]{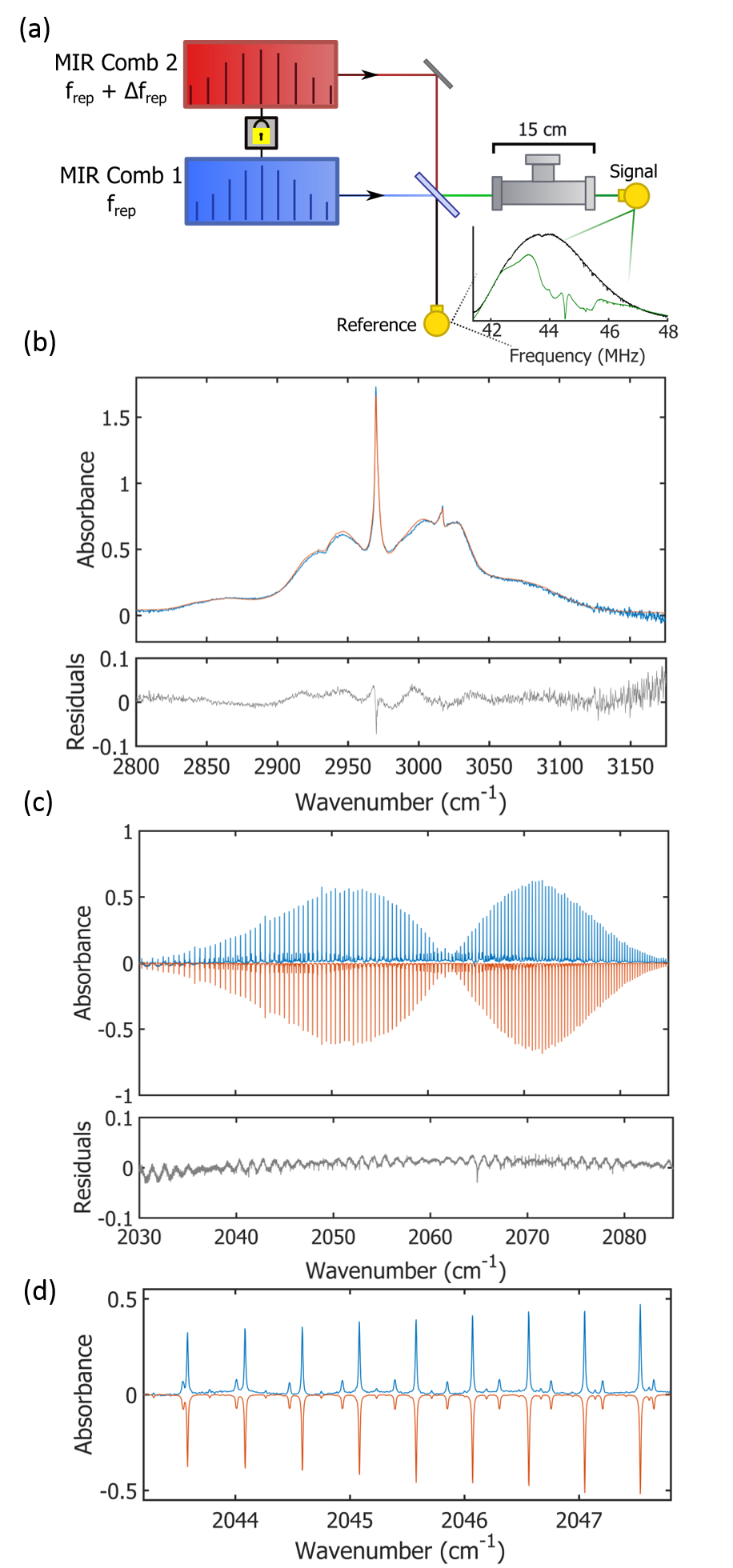}
\caption{Dual-Comb Spectroscopy. (a) Two MIR combs with slightly different repetition rates (50 Hz) combine on a 50/50 CaF beamsplitter. One combined beam is directly detected on a MCT detector to provide a reference spectrum. The other beam is sent through a 15 cm-long gas cell and detected on a second MCT detector. The Fourier transform of the time-domain signal is shown in the inset. (b) Acetone dual comb spectroscopy. Good agreement between experimental (blue) and predicted (red) spectra over broad absorption features. (c)-(d) OCS dual comb spectroscopy. Due to a small difference in path length that is exposed to atmosphere between the two arms, there is additionally a small amount of uncompensated water absorption at 2065 $\text{cm}^{-1}$. The zoomed-in portion of the OCS data highlights good agreement over absorbances ranging 0.01 to 0.5 in only 90 seconds of averaging.}
\label{Fig:DCS}
\end{figure}

Utilizing the stable and tunable nature of the MIR light, we perform gas-phase dual-comb spectroscopy (DCS) \cite{coddington_dual-comb_2016} on acetone and carbonyl sulfide (OCS), which allows us to investigate both narrow- and broad-band absorbers. In DCS, two combs with overlapping optical spectra and slightly different repetition rates are combined onto a photodetector. Because of the offset in repetition rate, a multi-heterodyne beating with millions of comb modes occurs between adjacent modes of the two combs, which translates the overlap in optical spectra into the RF-domain (Fig \ref{Fig:DCS}(a)). Due to precise control and tunability of the offset frequency, neither the $f_0$ tones nor copies of the DCS signal, arising from the presence of additional combs containing offset frequencies, interfere with the dual comb signal in the RF domain.

We generate the second MIR comb by duplicating the short pulse synthesis with a second Er:fiber oscillator. We then pump a second PPLN crystal with the same poling period, generating a similar second MIR comb. We combine the two MIR beams on a 50/50 CaF$_2$ beamsplitter and send both arms to separate MCT detectors (Fig \ref{Fig:DCS}(a)). In one of the combined beams, we place a 15 cm-long gas cell capped with wedged Si windows. The other path provides a simultaneous reference spectrum. The difference in repetition rates between the two oscillators is 50 Hz, which results in 20 ms-long interferograms.

To demonstrate the spectral stability and excellent baseline of the MIR combs, we first investigate the infrared spectrum of acetone, a broadband absorber (Fig \ref{Fig:DCS}(b)). We fill the gas cell by placing a few drops of liquid acetone into the cell and sealing it. After a few moments, the liquid evaporates, such that the cell is filled with acetone vapor in atmospheric background, significantly broadening the absorption features. Due to this broadening, we apodize the time-domain interferogram to 200 \textmu s, which brings our effective resolution to 0.3 $\text{cm}^{-1}$ (10 GHz). We average interferogram acquisitions over approximately 5 minutes. These data only require a linear baseline correction to agree with the model, provided by the Pacific Northwest National Laboratory (PNNL) Infrared Database \cite{sharpe_gas-phase_2004}. The additional noise on the blue side of the spectrum is due to a power drop off in the optical spectrum, reducing the signal-to-noise in this region.

We also investigate the infrared spectrum of OCS, highlighting the resolution achievable with these combs (Fig \ref{Fig:DCS}(c)\textendash{}(d)). For this measurement, we filled the gas cell with a mixture of gases, in which the partial pressure of OCS is $\approx$ 67 \textmu bar, with a background pressure of $\approx$ 67 mbar. Maintaining a low background pressure in the cell allows us to look at narrow absorption features less impacted by pressure broadening. For these data, we use the full 20 ms interferogram, providing a resolution of 0.003 $\text{cm}^{-1}$ (100 MHz), while only averaging over 1.5 minutes of data, still with only a linear baseline correction. The model comes from the HITRAN database \cite{rothman_hitran2012_2013}. We are able to accurately reconstruct the spectral envelope (Fig \ref{Fig:DCS}(c)), as well as match individual lines in both magnitude and line center (Fig \ref{Fig:DCS}(d)). The experimental data include a baseline oscillation due to an etalon in the path of the beam; however, we are still able to accurately reproduce each line in the model.

In this measurement, we estimate a peak frequency-domain SNR of 76 Hz$^{1/2}$. The high SNR here enables short averaging times for near real-time spectroscopic monitoring. This is competitive with other recent measurements in this wavelength region \cite{ycas_high-coherence_2018,spaun_continuous_2016,muraviev_massively_2018}, especially when accounting for the large optical bandwidths and small footprint (\textless 0.25 m$^2$) achieved here in contrast to prior work. 


We have presented a simple and robust method of generating optical frequency combs from 3 \textmu m to 5.5 \textmu m by pumping a PPLN crystal with an ultrashort pulse utilizing mature erbium fiber technology. We can measure offset frequency interference beats throughout the generated spectrum and use the beat measured directly from the MCT detector in stabilizing the combs to an in-loop timing jitter \textless 100 as.

Looking forward, the periodic poling can be engineered to generate combs covering the entire MIR atmospheric window, which would provide simultaneous access to carbon dioxide, various alcohols, carbonyl sulfide, methane, and many other atmospheric components and pollutants. Moreover, direct measurements of the mid-infrared electric field could help further elucidate the origin of the offset frequency beat in that spectral region.


\noindent\textbf{Funding} NIST Greenhouse Gas and Climate Science Measurement Program; the DARPA SCOUT program; National Science Foundation (NSF) (1708743); Air Force Office of Scientific Research (AFOSR) (FA9550-16- 1-0016). 

\noindent\textbf{Acknowledgements} This work is a contribution of the United States government and is not subject to copyright in the United States of America.

\bibliographystyle{apsrev4-1}
\bibliography{Zotero}

\end{document}